\documentclass[twocolumn,prb,aps,showpacs]{revtex4}

\usepackage{dcolumn}
\usepackage{amsmath}
\usepackage{epsfig}

\preprint{DRAFT}

\begin{document}


\title{Charge and orbital excitations in Li$_2$CuO$_2$}
\author{Young-June Kim}
\author{J. P. Hill}
\affiliation{Department of Physics, Brookhaven National Laboratory,
Upton, New York 11973}
\author{F. C. Chou}
\affiliation{Center for Materials Science and Engineering,
Massacusetts Institute of Technology, Cambridge, Massachusetts
02139}
\author{D. Casa}
\author{T. Gog}
\author{C. T. Venkataraman}
\affiliation{CMC-CAT, Advanced Photon Source, Argonne National
Laboratory, Argonne, Illinois 60439}

\date{\today}

\begin{abstract}

We report a resonant inelastic x-ray scattering study of electronic
excitations in Li$_2$CuO$_2$, an insulating compound comprised of ribbons of
edge-sharing copper-oxygen chains. Three excitations, which show little
dependence on momentum transfer, are observed in our measurements. The lowest
energy excitation at $\sim 2.1$ eV is dispersionless and is attributed to a
localized $d-d$ orbital excitation. We also observe two excitations at $\sim
5.4$ eV and $\sim 7.6$ eV which we assign to charge-transfer excitations.
These high-energy excitations are also dispersionless along the copper-oxygen
chain direction. However, in each case we observe a small energy dispersion
along the direction perpendicular to the copper-oxygen ribbons, suggesting a
significant interchain coupling in this system. We also discuss the possible
implications of ferromagnetic nearest-neighbor intrachain coupling on the
charge excitation spectra.

\end{abstract}


\maketitle

\section{Introduction}

The magnetic properties of insulating copper oxide compounds have been drawn
much attention over the past decade. Such systems include
quasi-one-dimensional (1D) spin chains (Sr$_2$CuO$_3$, SrCuO$_2$, CuGeO$_3$),
spin ladders (SrCu$_2$O$_3$, Sr$_{14}$Cu$_{24}$O$_{41}$), and
quasi-two-dimensional (2D) parent compounds of high temperature
superconductors (La$_2$CuO$_4$, Sr$_2$CuO$_2$Cl$_2$). Many of these systems
can be modeled by a simple Heisenberg spin Hamiltonian with only one
parameter, that is, the superexchange coupling between copper spins, $J$.
Unfortunately, calculating $J$ is a difficult task, due to the strong
electron correlations in these so-called Mott insulators. Part of the
difficulty also lies in the fact that there is limited information on the
electronic structure of these copper oxide compounds. Thus, spectroscopic
studies of the electronic structure and excitations can provide important
information leading towards an improved microscopic understanding of
magnetism in insulating copper oxides.

Li$_2$CuO$_2$ has been frequently modeled as an edge-sharing chain
compound,\cite{Hoppe70,Sapina90,Ohta93,Mizuno98,Boehm98,Neudert99,Atzkern00}
that is, the copper-oxygen plaquettes in this material are connected by their
edges with the Cu-O-Cu bond angle ($\theta$) close to 90$^\circ$.  
Electronic properties of Li$_2$CuO$_2$ have been studied with optical
conductivity,\cite{Mizuno98} x-ray absorption spectroscopy
(XAS),\cite{Neudert99} and electron energy-loss spectroscopy
(EELS).\cite{Atzkern00} The consensus from these experiments is that
Li$_2$CuO$_2$ is a charge-transfer (CT) insulator, with a CT gap of $2.2 \sim
2.7$ eV. This is in agreement with spin-polarized local density approximation
(LDA) calculations by Weht and Pickett,\cite{Weht98} which found a gap of 2.5
eV. Another important observation is that there exist several exchange paths
between the copper spins and that none of these exchange interactions
dominate, and as a result Li$_2$CuO$_2$ cannot be described as a simple
Heisenberg spin chain.\cite{Neudert99,Atzkern00} This intricate nature of the
magnetic interactions is most clearly demonstrated by the magnetic phase
transitions: Experimentally, the paramagnetic susceptibility of Li$_2$CuO$_2$
exhibits an antiferromagnetic Curie-Weiss behavior, with a transition
temperature $T_N \approx 9$ K. However, the magnetic structure is composed of
ferromagnetic chains, which are coupled antiferromagnetically. In addition, a
second transition to canted ferromagnetic phase at $\approx 2.8$ K has been
observed by magnetization\cite{Ortega98} and muon-spin rotation
studies.\cite{Staub00}

Due to this complexity of the underlying spin Hamiltonian, it is difficult to
experimentally determine the magnetic interactions in this system. Even the
sign of $J$, that is, whether the nearest-neighbor coupling along the chain
is ferromagnetic or antiferromagnetic is controversial. According to the
Goodenough-Kanamori-Anderson rules, $J$ is expected to be small, unlike the
large antiferromagnetic superexchange coupling present in the case of a
$\theta=180^\circ$ bond. In fact, calculations of Mizuno and
coworkers\cite{Mizuno98} show that $J$ is very sensitive to $\theta$ for
$\theta$ close to $90^\circ$. In the case of La$_6$Ca$_{8}$Cu$_{24}$O$_{41}$,
$\theta=91^\circ$ and $J$ is ferromagnetic ($J<0$), while for
$\theta=99^\circ$ in CuGeO$_3$, $J>0$. For Li$_2$CuO$_2$, a powder x-ray
diffraction study found $\theta=94^\circ$.\cite{Hoppe70} A number of
theoretical studies\cite{Mizuno98,Mizuno99,deGraaf02} suggest a ferromagnetic
coupling between nearest-neighbor copper spins, which is consistent with the
magnetic structure determined by powder neutron diffraction.\cite{Sapina90}
On the other hand, Boehm et al.  \cite{Boehm98} found that $J$ is
antiferromagnetic from their inelastic neutron scattering experiment.
Clearly, further investigation of the electron hopping and the exchange
interactions in this system is needed to determine the spin Hamiltonian and
elucidate the magnetic phase behavior.

In the present work, we have carried out resonant inelastic x-ray scattering
(RIXS) experiments to study the electronic excitations of Li$_2$CuO$_2$.  
This technique allows one to probe the energy and momentum dependence of
charge neutral electronic excitations, in particular focusing on those that
involve the copper orbitals.\cite{Hill98,Abbamonte99,Hasan00,Kim02a} We have
observed two types of excitations;  one at 2.1 eV which we attribute to a
localized $d-d$ type orbital excitation, and excitations at 5.4 eV and 7.6 eV
which we believe arise from CT-type processes in the Cu-O plaquettes. A
small, but finite, dispersion of the latter along the direction perpendicular
to the chain direction suggests that the interchain coupling is non-zero,
supporting the conclusions of previous studies of magnetic
interactions.\cite{Boehm98,Neudert99,Mizuno99,deGraaf02} In addition, we
compare our results with those of CuGeO$_3$ and discuss the possible
implications of ferromagnetic $J$ on our RIXS spectra.

In the next section, we describe the experimental configurations used in the
measurements. The incident energy dependence and the momentum dependence of
the observed RIXS spectra are discussed in Sec.~\ref{sec:resulta} and
Sec.~\ref{sec:resultb}, respectively. Finally, we will discuss the possible
implications of the experimental results in Sec.~\ref{sec:discussion}.

\section{experimental details}

The experiments were carried out at the Advanced Photon Source on the
undulator beamline 9ID-B. A double-bounce Si(333) monochromator and a
spherical, diced, Ge(733) analyzer was used to obtain an overall energy
resolution of 0.4 eV (FWHM). The scattering plane was vertical and the
polarization of the incident x-ray was kept close to the {\bf c}-direction
for the data reported here. Note that the edge-sharing CuO$_4$ chain runs
along the crystallographic {\bf b}-direction, and the Cu-O plaquettes lie
perpendicular to the {\bf a}-direction.\cite{Hoppe70} A single crystal
sample of Li$_2$CuO$_2$ ($a=3.662$ \AA, $b=2.863$ \AA, and $c=9.393$ \AA)
was grown using the traveling solvent floating zone method. The crystal
was cleaved along the $(1 \; 0 \; 1)$ plane and mounted on an aluminum 
sample holder at room temperature. Since Li$_2$CuO$_2$ is 
hygroscopic, care was taken to cleave the sample immediately before the 
experiment. It was then kept in vacuum throughout the experiment.

\section{experimental results}

\subsection{Incident Energy Dependence}
\label{sec:resulta}

\begin{figure} \begin{center}
\epsfig{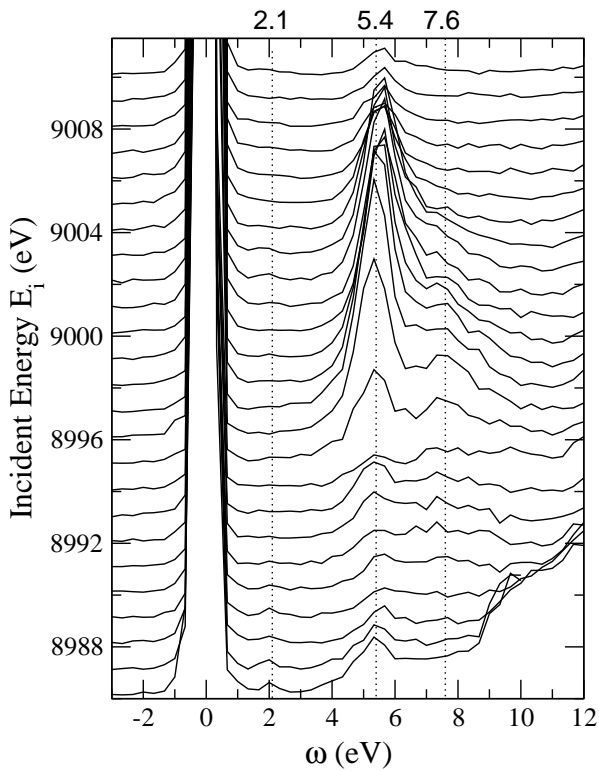}
\end{center} 
\caption{ Scattered intensity at {\bf Q}=(2.5 0.1 
2.5) as a function of energy transfer, $\omega$. For clarity, scans are 
shifted vertically, and 
the incident energy for
each scan can be read off from the vertical axis. The incident polarization 
was along the {\bf c}-direction (i.e., in the plane of the CuO$_4$ 
plaquettes). Error bars are omitted for clarity.}
\label{fig1} 
\end{figure}

In Fig.\ \ref{fig1}, we plot the incident energy ($E_i$) dependence of the
scattered intensity as a function of energy transfer ($\omega$) at a fixed
momentum transfer of {\bf Q}={(2.5 0.1 2.5)}.  Since {(2.5 0.1 2.5)} is near
the Brillouin zone boundary, the elastic scattering intenisty (i.e., at
$\omega=0$)  is not very large. Three resonant features are observed in Fig.\
\ref{fig1}. The strongest feature has an excitation energy of $\sim 5.4$ eV,
and shows a large resonant enhancement in the intensity as the incident
energy is varied through $E_i \approx 9000$ eV, becoming weaker as the
incident energy is tuned away from the resonance. A second feature at $\sim
7.6$ eV shows slightly different resonance behavior. In addition, a third
feature at 2.1 eV is very weak compared to the other two features, and
resonates at much lower incident energy, around $E_i \lesssim 8990 $ eV.
Note, because of geometrical constraints associated with the near
backscattering angle of the Ge(733)  analyzer crystal at these low incident
energies, it was not possible to scan the incident energy below $E_i=8979$
eV.

\begin{figure} \begin{center}
\epsfig{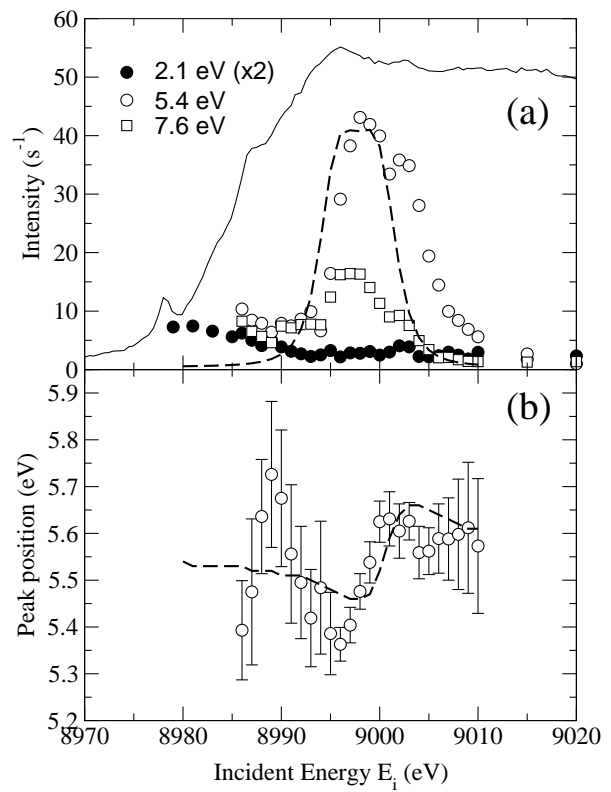}
\end{center} 
\caption{
(a) Scattered intensity at the three energy transfers, 2.1 eV, 5.4 eV, and
7.6 eV, as a function of $E_i$ (raw data). Also plotted is the x-ray
absorption, as measured by monitoring the fluorescence yield from the sample
in the same scattering geometry (solid line). (b) The fitted peak position of
the 5.4 eV feature as a function of $E_i$. The dashed lines in both figures
are the results from fitting the data of Fig.~\ref{fig1} to Eq.~(\ref{eq1}),
as described in the text.
}
\label{fig2} 
\end{figure}

To illustrate the resonance profile of these peaks, we plot in
Fig.~\ref{fig2}(a) the scattered intensity as a function of $E_i$, with the
energy transfer fixed at these excitation energies.  That is, the
intensity as measured at the positions indicated by vertical dotted lines in
Fig.~\ref{fig1} is plotted as a function of $E_i$ in Fig.~\ref{fig2}(a). The
intensity of the 2.1 eV feature is multiplied by 2 in order to show its
resonance profile more clearly. Raw experimental data are plotted
without any absorption corrections in this figure. For the 5.4 eV feature,
the observed spectra were also fitted to a Lorentzian squared lineshape. The
fitted intensity as a function of $E_i$ (not shown) is virtually identical to
the raw data shown in Fig.~\ref{fig2}(a). However, the peak position of the
5.4 eV feature, plotted in Fig.~\ref{fig2}(b), exhibits an interesting
$E_i$-dependence. Specifically, there is a linear shift of the peak position
from 5.4 eV at $E_i = 8995$ eV to 5.6 eV at $E_i = 9000$ eV; for $E_i \gtrsim
9000$ eV, the peak position remains around 5.6 eV. An apparent peak position
shift is also observed for $E_i \lesssim 8995 $ eV. However, the small
intensity [shown in Fig.~\ref{fig2}(a)] and the resultant large error bars
for these data points make it difficult to analyze in quantitatively
meaningful way. Thus, we will focus on the peak position shift for $E_i
\gtrsim 8995$ eV in what follows.

Such an apparent shift at first sight appears surprising. The natural
expectation is that a valence excitation -- such as this -- would appear as a
Raman-shift in a RIXS spectra, that is, a peak at a fixed energy-transfer,
independent of the incident energy. However, in their RIXS study of
La$_2$CuO$_4$, Abbamonte and coworkers also observed an incident energy
dependence of the peak position.\cite{Abbamonte99} They proposed an
expression based on a shakeup picture in third order perturbation theory to
explain the experimental results, following earlier theoretical work by
Platzman and Isaacs.\cite{Platzman98} Specifically, the scattered intensity
$w$ is expected to have the form
\begin{equation} 
w =
{S_K({\bf q},\omega)  \over{[(E_i-E_K)^2+\gamma_K^2][(E_f-E_K)^2+\gamma_K^2]}}, \label{eq1}
\end{equation} 
where $E_K$ and $\gamma_K$ are adjustable parameters, and $S_K$ describes an
electronic excitation spectrum of interest. The key ingredient of
Eq.~(\ref{eq1}) is that both incoming and outgoing resonances are included in
the denominator.  The shape of this function depends on the relative widths
of the function $S_K({\bf q},\omega)$ and the inverse lifetime, $\gamma_K$.
For a sharp $S_K({\bf q},\omega)$, $\omega$ is peaked at constant energy
transfer as a function of $E_i$. For a small value of $\gamma_K$, the peak
exhibits characteristic dispersion. We have fitted all the data in the range
of $3 < \omega < 7$ from Fig.~\ref{fig1} to Eq.~(\ref{eq1}), using a
Lorentzian squared function for $S_K$ of full-width 1.1 eV. We have obtained
a single set of parameters $E_K \approx 8995$ eV and $\gamma_K \approx 2.5$
eV. These parameter values are similar to those obtained in
Ref.~\onlinecite{Abbamonte99}. By substituting the values of these parameters
back in Eq.~(\ref{eq1}), we can reproduce the intensity and the peak
position, which is plotted as a dashed line in Fig.~\ref{fig2}. The fitting
results indeed describe the observed peak shift and intensity on a {\it
qualitative} level.  However, it is not at all clear that this is a unique
description of the data and further understanding of the discrepancy between
the fit and the observed data will require a detailed model calculation as
well as a microscopic theory of RIXS cross-section. We note that in the soft
x-ray regime, there exist detailed model calculations to describe resonant
emission spectroscopy experiments.\cite{Mizouchi98,Okada00}

Also shown in Fig.~\ref{fig2}(b) is the measured x-ray absorption spectrum
(solid line).  The final states of this process are the
intermediate states of the RIXS process. One can associate the peaks
around 8986 eV and 8996 eV in the absorption spectrum with the
well-screened ($\underline{1s} 3d^{10} \underline{L} 4p$) and
poorly-screened ($\underline{1s} 3d^9 4p$) core hole final states,
respectively, where $\underline{1s}$ and $\underline{L}$ denote holes in
the core level and oxygen ligands, respectively. This association is
consistent with the XAS studies on other insulating cuprates, including
CuGeO$_3$. \cite{Cruz99} We also observe a small pre-edge feature around
$E_i=8978$ eV. Qualitatively similar results were obtained in the XAS
study of CuGeO$_3$ by Cruz et al.,\cite{Cruz99} in which the pre-edge
feature was assigned to electric dipole forbidden transitions from the Cu
$1s$ level to unoccupied Cu orbitals with $3d$ character.  The possible
channels suggested for such transitions were (a) electric dipole
transitions to Cu $4p$ character mixed with neighboring Cu $3d$ states,
and (b) electric quadrupole transitions to the Cu $3d$ states. In
either case, the pre-edge feature, which is the intermediate state for the
peak at 2.1 eV, has a strong $3d$ character and will therefore have a
large overlap with excitations involving $3d$ electrons.

The intermediate states responsible for the resonant enhancement of the
high-energy features at 5.4 eV and 7.6 eV are the poorly-screened states.
This is consistent with previous RIXS results obtained for quasi-2D cuprates
La$_2$CuO$_4$ \cite{Kim02a} or Nd$_2$CuO$_4$.\cite{Hill98} However, the
association of RIXS excitations with a particular intermediate states is not
absolute. In Fig.~\ref{fig1}, for example, the 5.4 eV feature seems to
resonate at more than one intermediate state; that is, the intensity
increases again as the incident energy decreases below $\sim 8990$ eV. This
needs further investigation.

\begin{figure}
\begin{center}
\epsfig{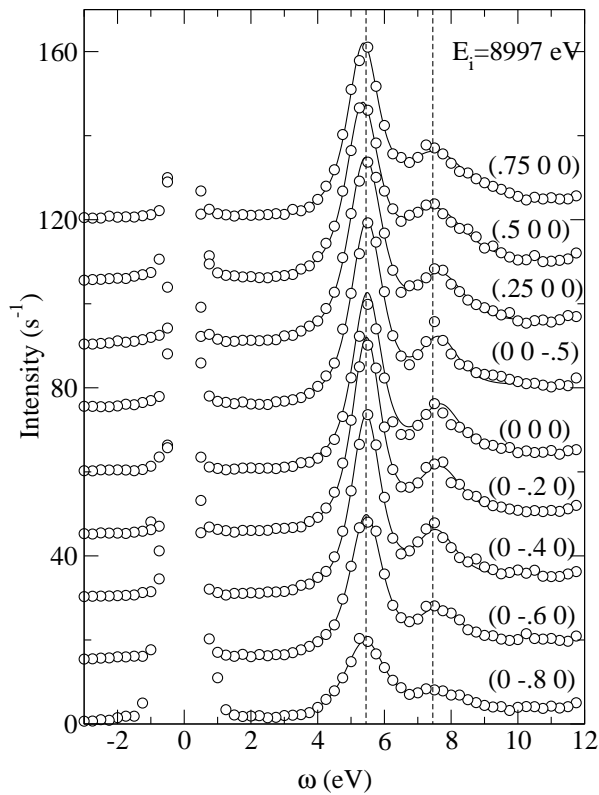}
\end{center}
\caption{
Energy loss spectra taken at fixed $E_i=8997$ eV at a reduced
wave vector [${\bf q} \equiv {\bf Q} - (2 0 3)$] as noted. 
Solid lines through the data points are the fit results as 
described in the text. Dashes lines show that the peak positions have very 
little {\bf q}-dependence. Each spectrum is offset
vertically for clarity. Error bars are smaller than the symbol size.
}
\label{fig3}
\end{figure}

\subsection{Momentum Dependence}
\label{sec:resultb}

\begin{figure}
\begin{center}
\epsfig{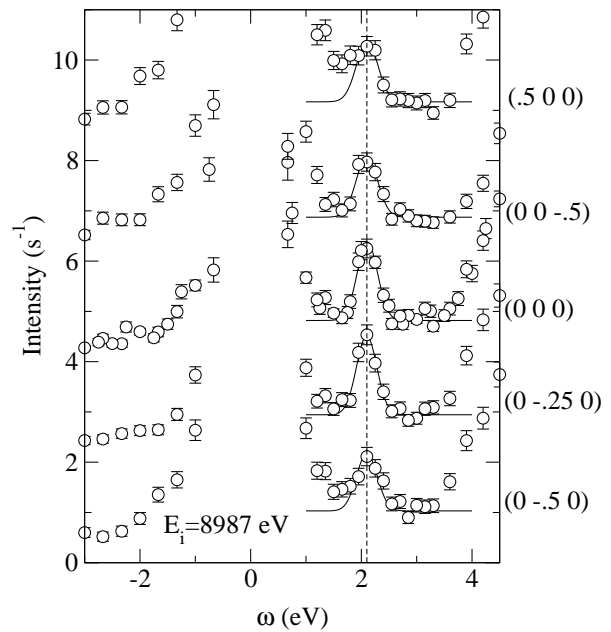}
\end{center}
\caption{
Energy loss spectra for the 2.1 eV feature is shown with $E_i=8987$ eV for 
a fixed reduced
wave vector ({\bf q}) as noted. Solid lines through data points are 
fitting results as described in the text.
Dashes lines show that the peak positions have very little {\bf 
q}-dependence. 
Each spectrum is offset
vertically for clarity.
}
\label{fig5}
\end{figure}

\begin{figure}
\begin{center}
\epsfig{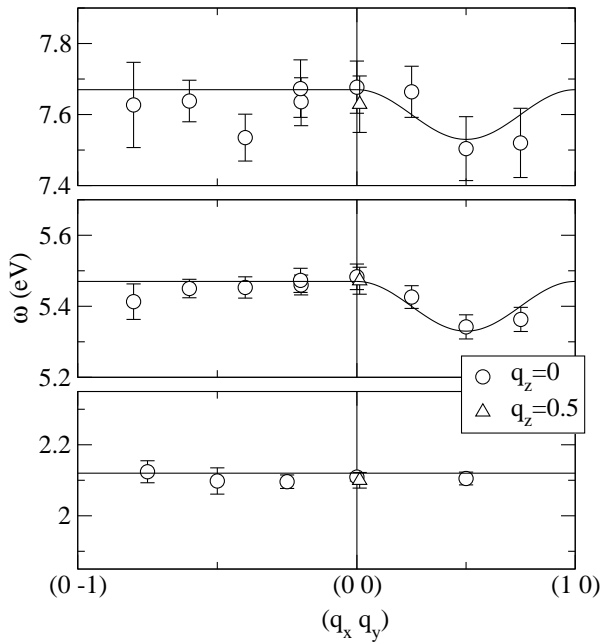}
\end{center}   
\caption{
The peak positions of the three features obtained 
from fitting scans shown in Figs. 2 and 3 to a Lorentzian-squared 
lineshape (5.4 eV and 7.6 eV) or a Gaussian lineshape (2.1 eV).
Data on the left panels are taken at the position with ${\bf 
q}\parallel(100)$, and those on the right panels with ${\bf
q}\parallel(010)$. The data taken at {\bf q}=(0 0 -0.5) are shown as 
triangles.
}
\label{fig4}
\end{figure}

To investigate the momentum dependence of these excitations, we have measured
the energy-loss spectra along the high-symmetry directions [100], [010], and
[001], around the (2 0 3) reciprocal lattice point. In Figs.~\ref{fig3} and
\ref{fig5}, energy-loss scans taken with the incident energy fixed at
$E_i$=8997 eV and $E_i$=8987 eV, respectively, are plotted at various
momentum transfers. The reduced wave vector, {\bf q}, is noted for each scan.
Vertical dashed lines are drawn to show the almost dispersionless behavior of
the peaks at 5.4 eV and 7.6 eV in Fig.~\ref{fig3}, and at 2.1 eV in
Fig.~\ref{fig5}.  To obtain quantitative information on the dispersion of
these excitations in Fig.~\ref{fig3}, we fit the observed spectra with two
peaks; both with a Lorentzian squared lineshape. We note here that these
peaks are not resolution limited; the full-widths of the 5.4 eV peak and the
7.6 eV peak were fixed at 1.1 eV and 1.5 eV, respectively. The peak positions
obtained from these fits are plotted in the upper two panels of
Fig.~\ref{fig4}. Due to its proximity to the much stronger peak at 5.4 eV,
the location of the peak at 7.6 eV has rather large error bars, and more or
less follows the dispersion of the 5.4 eV feature. As evident in
Fig.~\ref{fig3}, the energy dispersions of both excitations are very small
along all high-symmetry directions. In particular, along the chain direction,
[0 1 0], the 5.4 eV feature is virtually dispersionless. On the other hand,
there is a suggestion of some dispersion ($\sim 140$ meV) along the [1 0 0]
direction, which is perpendicular to the copper-oxygen ribbon.

We attribute the 5.4 eV feature to a charge-transfer type excitation on a
single copper-oxygen plaquette; specifically, we believe that it represents
the energy difference between a bonding state and an antibonding state. Due
to the strong hybridization between the Cu $3d$ and O $2p$ orbitals, the
ground state for a half-filled copper-oxygen plane is not a simple 3$d^9$.
Rather, according to the Anderson impurity model, \cite{Hill98} it is a
bonding state with an admixture of 3$d^9$ state and $3d^{10}\underline{L}$
state.  In the RIXS process at the Cu K-edge, a Cu $1s$ electron is excited
to the Cu $4p$ band, and this intermediate state may then decay into the
antibonding excited state, producing the 6 eV feature commonly found in
cuprate systems.\cite{Hill98,Abbamonte99} This interpretation is also
consistent with the calculation by Weht and Pickett,\cite{Weht98} in which
they found a splitting of $\sim 5$ eV between the bonding and the antibonding
states. Since this excitation is localized in a single plaquette of one
copper and four oxygens, its excitation energy is somewhat material
independent and would also be expected to have a very small momentum
dependence. Recently, a small ($\sim 150$ meV)  dispersion along the chain
direction has been observed for the 6.4 eV excitation of
CuGeO$_3$.\cite{Zimmermann02} Since this dispersion should depend critically
on the angle $\theta$ formed by Cu-O-Cu bond,\cite{Mizuno98} it is not
surprising to observe smaller dispersion in Li$_2$CuO$_2$ ($\theta=94^\circ$)
than in CuGeO$_3$ ($\theta=99^\circ$).

There are at least two possible interpretations for the 7.6 eV feature.  
First, the peak at 7.6 eV could arise from excitation of electrons from
different bands than for the 5.4 eV feature.  Alternatively, in the localized
excitation picture described above, and in Ref.~\onlinecite{Atzkern00}, one
can have different excitation modes depending on the symmetry of the four
oxygen orbitals in the plaquette. In this scenario, the 7.6 eV feature would
involve the same bands, but differ from the 5.4 eV feature by the symmetry of
the particle-hole pair. Further studies of polarization dependences might be
able to distinguish between these two possibilities.

We next discuss the momentum dependence of the 2.1 eV feature. In order to
investigate this, the incident photon energy was held fixed at $E_i$=8987 eV
and energy loss scans were taken at a number of momentum transfers
(Fig.~\ref{fig5}). In contrast to the higher energy features, the 2.1 eV
feature is found to be resolution limited at all momentum transfers and is
therefore fitted to a single Gaussian lineshape. The peak position extracted in
this way is plotted in the bottom panel of Fig.~\ref{fig4}.  It is clear that
the energy dispersion of this peak is negligibly small. The two salient
characteristics of the 2.1 eV feature are thus its lack of dispersion and its
narrow peak width, which implies that this excitation is localized and has a
long-lifetime. Another important point is that the intermediate states of this
2.1 eV feature are either $3d$ states, or states having a large overlap with
the $3d$ states.

Based on these observations, we associate the 2.1 eV feature with a $d-d$
type orbital excitation; that is, an excitation corresponding to exciting
holes from the $d_{yz}$ orbital to higher energy $d$-orbitals. According to
the calculation of Tanaka et al.,\cite{Tanaka99} the energy splitting between
the ground state, $d_{yz}$, and excited levels of $d_{xy}$, $d_{zx}$,
$d_{3x^2-r^2}$ in Li$_2$CuO$_2$ is $\sim 2$ eV, which is consistent with our
value 2.1 eV. Similar $d-d$ excitations have been observed in CuGeO$_3$ with
RIXS \cite{Zimmermann02} and optical spectroscopies.\cite{Bassi96}

\section{discussion}
\label{sec:discussion}

To summarize, we have observed energy loss features at 2.1 eV, which we
attribute to a $d-d$ excitation, and at 5.4 eV and 7.6 eV, which are
attributed to localized charge-transfer excitations of copper oxygen
plaquettes. These obsevations are generally consistent with the recent RIXS
study of a similar edge-sharing chain compound CuGeO$_3$, with only a
quantitative difference in the excitation energies.  However, an additional
excitation at 3.8 eV was observed for CuGeO$_3$. This was assigned to a
non-local CT exciton mode formed by a particle and hole pair residing on
neighboring plaquettes. The particle in this case forms a $d^{10}$ state on
one plaquette and the hole forms a Zhang-Rice (ZR) singlet state
($d^{9}\underline{L}$) on the neighboring plaquettes.\cite{Zhang88} The
exciton state formed by this particle-hole pair can have a large dispersion
of $\sim 1$ eV in corner-sharing geometries such as
La$_2$CuO$_4$,\cite{Kim02a} Sr$_2$CuO$_3$,\cite{Hasan02} or
SrCuO$_2$.\cite{Kim03b} In these cuprate materials (including CuGeO$_3$), the
RIXS peak corresponding to this CT exciton was observed at energies slightly
higher than the CT gap energy as measured in optical conductivity
[$\sigma(\omega)$] studies. This trend is illustrated in Table~\ref{table1},
where we list the RIXS peak positions and the corresponding optical
conductivity peak positions in selected cuprate compounds. This difference
presumably originates in the difference in the measured response functions.
That is, the optical conductivity is proportional to the imaginary part of
complex dielectric function, Im$[\epsilon({\bf q}=0,\omega)]$, while the RIXS
cross section is expected to follow the dielectric loss function,
Im$(-1/\epsilon({\bf q},\omega)$.

\begin{table}
\caption
{RIXS peak position of CT exciton and corresponding peak position 
in optical conductivity $\sigma(\omega)$.}
\label{table1}
\begin{ruledtabular}
\begin{tabular}{lcccr}
 & RIXS (eV) & Ref. & $\sigma(\omega)$ (eV) &Ref.\\
\hline
$\rm CuGeO_3$&          3.8&   \onlinecite{Zimmermann02}&   3.4&   
\onlinecite{Damascelli00}\\
$\rm La_2CuO_4$&        2.2&   \onlinecite{Kim02a}&   2.0&   
\onlinecite{Uchida91}\\   
$\rm SrCuO_2$&          2.5&   \onlinecite{Kim03b}&   1.8&   
\onlinecite{Popovic01}
\end{tabular}
\end{ruledtabular}
\end{table}

In light of the above discussion, it seems clear that none of the observed
features in Li$_2$CuO$_2$ are such a non-local, exciton-like excitation.
Consider first the case of the 2.1 eV feature, which we have previously
argued as a $d-d$ excitation. Were this excitation to be in fact, a non-local
CT excitation, then, as discussed above, one would expect to observe a sharp
feature in the optical conductivity just below this value, say around 1.9 eV.  
No such feature is observed in the optical data.\cite{Mizuno98} Rather, one
observes a rapid increase of the optical conductivity above 2.6 eV which
develops into peaks around 3.5 eV and 4.2 eV. In addition, as noted, the 2.1
eV feature is resolution limited, while CT exciton features are characterized
by broad peaks in other materials.\cite{Kim02a,Zimmermann02,Kim03b}
Conversely, the feature at 5.4 eV is too high in energy to be associated with
such a non-local excitation, both in terms of theoretical expectations for
such an excitation and from the observed features in the optical conductivity
data.  Thus we conclude that such non-local CT exciton-like excitations are
suppressed in Li$_2$CuO$_2$.

This immediately raises the question as to why such excitations are
suppressed in this material. One possible explanation for this apparent
absence lies in the different magnetic ground states of CuGeO$_3$ and
Li$_2$CuO$_2$.  This non-local CT exciton involves the movement of the hole
on one copper site onto the oxygen orbitals of the neighboring plaquette
where it forms a singlet state with the copper spin on that plaquette. This
hole necessarily preserves the spin of the original Cu $3d$ hole. If
neighboring copper spins are ferromagnetically coupled, as is the case for
Li$_2$CuO$_2$, then a singlet state cannot be formed and only triplet
excitations are possible. Since a significant fraction of the exciton
formation energy comes from the binding energy of the Zhang-Rice
singlet,\cite{Zhang98} this excitation is likely to be significantly
suppressed in such cases. Conversely, for the case of antiferromagnetic
coupling of neighboring coppers, as in the case of CuGeO$_3$, the singlet
state may form naturally and the non-local excitation is stabilized. This
argument was first made in the context of O K-edge RIXS spectra by Okada and
Kotani\cite{Okada01} who showed just such an effect theoretically in
calculated spectra for edge-sharing geometries, and it seems plausible that
it is also active in the present case.  The absence of such an exciton would
be consistent with the absence of a sharp feature near the CT gap in the
optical conductivity data, and in fact, a recent theoretical calculation on a
small cluster seems to support these ideas.\cite{Maekawa} In any case, it is
quite remarkable that the sign of the nearest-neighbor exchange coupling
manifests itself at room temperature via the (non)existence of the RIXS peak
from the CT exciton. This further emphasizes the delicate nature of bond
angles and orbital overlap in determining the electronic properties of these
strongly correlated cuprate systems.

Finally, in order to describe the apparent dispersion along the {\bf
a}-direction, one needs to go beyond the simple picture of a localized
excitation. As shown in Fig.~\ref{fig5}, the features at 5.4 eV and 7.6 eV
seem to exhibit non-zero dispersion along the {\bf a}-direction, with a
minimum located at the zone boundary. This is similar to the dispersion
behavior of the 6.4 eV excitation of CuGeO$_3$, \cite{Zimmermann02} although
in that case dispersion was observed along the chain direction. If confirmed,
this observation suggests that the interchain coupling along the {\bf
a}-direction is not negligible.  However, according to the calculation
by de Graaf {\it et al.}.\cite{deGraaf02}, the superexchange interaction via
Li and O orbitals is too small to account for the non-zero dispersion
behavior along this direction. A different exchange mechanism, such as direct
exchange between Cu orbitals may need to be considered to understand
this.\cite{Neudert99}

In summary, we have observed three excitations in Li$_2$CuO$_2$ using resonant
inelastic x-ray scattering. These occur at 2.1, 5.4, and 7.6 eV and are
attributed to a $d-d$ excitation and two CT-type excitations, respectively.
None of the excitations exhibit measurable dispersion along the chain direction
consistent with the Cu-O-Cu bond angle being close to $90^\circ$. There is some
evidence of dispersion perpendicular to the plane of the Cu-O ribbon for the
CT-type excitations.  This would suggest there is a non-zero interchain
coupling. Finally, we find no evidence for a non-local (Zhang-Rice) CT exciton
in the vicinity of the gap. We associate the suppression of such a feature with
the ferromagnetic coupling between the neighboring copper spins.

\begin{acknowledgements}

We would like to thank S. Maekawa and J. van den Brink for invaluable
discussions. The work at Brookhaven was supported by the U. S. Department of
Energy, Division of Materials Science, under contract No. DE-AC02-98CH10886.
Use of the Advanced Photon Source was supported by the U. S. Department of
Energy, Basic Energy Sciences, Office of Science, under Contract No.  
W-31-109-Eng-38.

\end{acknowledgements}

\end{document}